\documentclass[prl,twocolumn,preprintnumbers]{revtex4}
\usepackage{amsmath,amssymb,epsfig}
\usepackage{times}
\usepackage{latexsym}
\usepackage{multirow}
\usepackage{url}
\usepackage{color}

\usepackage{graphicx}
\usepackage{epsfig}
\usepackage{amssymb}
\usepackage{color}

\newcommand{\bal}{\begin{aligned}} \newcommand{\eal}{\end{aligned}}

\newcommand{\comment}[1]{}

\DeclareMathOperator{\Div}{Div}

\def\IZ{\mathbb{Z}}

\def\cN{\mathcal{N}}
\def\cO{\mathcal{O}}
\def\cV{\mathcal{V}}

\def\cX{\mathcal{X}}

\def\e{\mathrm{e}}

\def\clap#1{\hbox to 0pt{\hss#1\hss}}

\def\mclap{\mathpalette\mathclapinternal}

\def\mathclapinternal#1#2{%
\clap{$\mathsurround=0pt#1{#2}$}}

\def\ce{\mathrel{\mathop:}=} 

\newcommand{\sect}[1]{\noindent\textit{#1}.---\!}
\renewcommand\tilde\widetilde

\begin{document}

\thispagestyle{empty}

\title{Calabi--Yau Manifolds with Large Volume Vacua}
\author{James Gray$^{a}$, Yang-Hui He$^{b}$, Vishnu Jejjala$^{c}$, Benjamin Jurke$^{d}$, Brent Nelson$^{d}$, Joan Sim\'on$^{e}$}
\affiliation{
(a) Arnold-Sommerfeld-Center for Theoretical Physics, Fakult\"at f\"ur Physik, Ludwig-Maximilians-Universit\"at M\"unchen, Theresienstra\ss e 37, 80333 M\"unchen, Germany \\
(b) Department of Mathematics, City University, London, Northampton Square, London EC1V 0HB, UK; \\
School of Physics, NanKai University, Tianjin, 300071, P.R.\ China; \\
and Merton College, University of Oxford, OX1 4JD, UK \\
(c) Centre for Theoretical Physics, NITheP, and School of Physics, University of the Witwatersrand, Johannesburg, WITS 2050, South Africa \\
(d) Department of Physics, Northeastern University, Boston, MA 02115, USA \\
(e) School of Mathematics and Maxwell Institute for Mathematical Sciences, King's Buildings, University of Edinburgh, Edinburgh, EH9 3JZ, UK
}

\begin{abstract}
We describe an efficient, construction independent, algorithmic test to determine whether Calabi--Yau threefolds admit a structure compatible with the Large Volume moduli stabilization scenario of type~IIB superstring theory.
Using the algorithm, we scan complete intersection and toric hypersurface Calabi--Yau threefolds with $2 \leq h^{1,1} \le 4$ and deduce that $418$ among $4434$ manifolds have a Large Volume Limit with a single large four-cycle.
We describe major extensions to this survey, which are currently underway.
\end{abstract}

\maketitle

\renewcommand{\thepage}{\arabic{page}}
\setcounter{page}{1}
\def\thefootnote{\arabic{footnote}}
\setcounter{footnote}{0}

\sect{Introduction}
A realistic string model of low energy physics requires the moduli of the associated compactification to be stabilized.
The Large Volume Scenario (LVS)~\cite{Balasubramanian:2005zx} presents one of the most promising avenues to such a goal.
In this approach a combination of fluxes, non-perturbative effects, and $\alpha'$ as well as loop expansion corrections are employed to generate a stable vacuum which is well within the regime of validity of a supergravity description of the theory.
One particularly nice feature of the LVS is that it avoids the ``fine tuning'' of flux parameters required by other scenarios such as that of KKLT~\cite{Kachru:2003aw,Giddings:2001yu}.
The LVS instead balances non-perturbative and perturbative effects in a controlled manner by exploiting a situation where the overall volume $\cV$ of a smooth Calabi--Yau threefold $\cX$ is exponentially larger than the scale associated with four-cycles wrapped by certain E$3$-brane instantons.
Manifolds which are capable of supporting an appropriate structure of small and large cycles are termed ``Swiss Cheese.''

One starting point in the construction of a LVS string model is to take $\cX$ to be a smooth Calabi--Yau threefold where the overall volume $\cV$ takes a distinctive diagonal form in terms of a single ``large'' four-cycle and a number of ``small'' four-cycles:
\begin{equation}\label{eq_SwissCheeseScaling}
\cV \sim \tau_{\rm large}^{\frac{3}{2}} - \sum_a \tau_{a, {\rm small}}^{\frac{3}{2}} ~.
\end{equation}
More general possibilities are available~\cite{Cicoli:2008va}, however, and as such we shall refer to Calabi--Yau manifolds of the type \eqref{eq_SwissCheeseScaling} as ``Strong Cheese.''
With this example geometry, the majority of four-cycles that are wrapped by E$3$-brane instantons are small while the Calabi--Yau volume, which gets exponentially large, addresses phenomenologically important hierarchy questions.
Moreover, the existence of the large cycle proffers a flat potential for cosmological inflation.
Different numbers of large and small cycles are also possible and interesting to study, although it should be noted that the main theorem of~\cite{pmh} states that all but a maximum of $19$ K\"ahler parameters can always be interpreted as describing resolutions of singularities.

Despite the promising features of the LVS, there is a relative scarcity of explicit examples \cite{Cicoli:2008va,Blumenhagen:2008zz,Cicoli:2011it,Cicoli:2012vw}.
Although some outstanding work studying classes of appropriate manifolds can be found in~\cite{Cicoli:2011it,Cicoli:2012vw}, the density of the Swiss Cheese geometries within the standard constructions of Calabi--Yau threefolds is not known.
It is the aim of this letter to improve upon the situation by providing an algorithm that scans for Swiss Cheese manifolds in as general a manner as possible.
In particular, this algorithm is independent of the construction of the Calabi--Yau threefold, can yield definite negative as well as positive results, and produces more general examples than those of the form (\ref{eq_SwissCheeseScaling}).
The analysis is exact and analytic;
we can look for any number of large and small cycles.
To illustrate the use of this algorithm we present a scan over the complete intersection Calabi--Yau manifolds in products of projective spaces (CICYs) as well as Calabi--Yau manifolds realized as hypersurfaces in toric varieties, with $h^{1,1}(\cX) \leq 4$.
We will see that there are no Swiss Cheeses among the former type of geometries, while the latter yields a rich set of new examples.\\

\vspace{-0.4cm}

\sect{Swiss Cheese Calabi--Yau}
We begin with some geometric preliminaries.
Let $D_i\subset\cX$ be four-cycle divisors of $\cX$.
The triple intersection numbers $\kappa_{ijk}$ are defined with respect to the basis $\{[D_i]\}$ for $H^{1,1}(\cX;\IZ)\cong\Div(X)$.
The symplectic K\"ahler $(1,1)$-form $J$ is parameterized by $h^{1,1}(\cX)$ K\"ahler parameters $t^i$,
\begin{equation}\label{eq:CY3SymplecticJ}
J = \sum_{i=1}^{\mclap{h^{1,1}}} t^i [D_i] ~,
\end{equation}
which endows the K\"ahler parameters $t^i$ with a natural interpretation as two-cycle volumes.
Likewise, the holomorphic $(3,0)$-volume form $\Omega$ that specifies the complex structure depends upon $h^{2,1}(\cX)$ parameters.

The overall volume $\cV$ of $\cX$ is determined by the K\"ahler parameters and intersection numbers:
\begin{equation}\label{eq:InnerVolume}
\cV = \frac{1}{3!}\int_\cX J\wedge J \wedge J = \frac{1}{6} \sum_{i,j,k} \kappa_{ijk} t^i t^j t^k ~.
\end{equation}
In a similar fashion, the volumes of the four-cycles $D_i\subset\cX$ are related to the parameters $t^i$ by
\begin{equation}\label{eq_taudef}
\tau_i = \frac{\partial\cV}{\partial t^i} = \frac{1}{2!}\int_\cX [D_i]\wedge J \wedge J = \frac{1}{2}\sum_{j,k} \kappa_{ijk} t^j t^k ~.
\end{equation}

%

The generic Calabi--Yau threefold admitting a large volume vacuum has a number of ``small'' four-cycles, whose volumes remain finite in the large volume limit (LVL), where the threefold's volume $\cV$ and the volumes of the ``large'' four-cycles become parametrically large.
The criterion for $\cX$ to be compatible with the LVS is in~\cite{Cicoli:2008va}.
For the convenience of the reader we reproduce parts of this discussion.

Let $\tau_1,\ldots,\tau_{N_{\rm small}}$ remain small as $\tau_{N_{\rm small}+1},\ldots,\tau_{h^{1,1}(\cX)}\to\infty$, sending $\cV\to\infty$.
The low energy limit of type~IIB string theory in the LVS is a $d=4$, $\cN=1$ supergravity.
The scalar potential $V$, which is constructed from the superpotential and the K\"ahler potential, admits a set of non-super\-symmetric AdS minima at exponentially large volume located at $\cV\sim\e^{a_i \tau_i}$ for all small cycles $i=1,\dots,N_{\rm small}$ and parameters $a_i$ appearing in the superpotential if and only if $h^{2,1}(\cX) > h^{1,1}(\cX) \ge 2$ and each small cycle of volume $\tau_j$ behaves like a blow up mode resolving a point-like singularity.
The first of these conditions leads us to consider only Calabi--Yau threefolds with negative Euler characteristic.

The essential property of $\cX$ established in~\cite{Cicoli:2008va} is that the inverse K\"ahler metric for the small four-cycles associated to the volumes $\tau_\alpha$ exhibits non-generic scaling properties with respect to large cycles.
For example, diagonal components of the inverse K\"ahler metric do not have a leading term which scales with the second power of large divisor volumes but rather has the form
\begin{equation}\label{eq_invKaehlerDiag}
K^{-1}_{\alpha\alpha}\sim\cV\sqrt{\tau_\alpha} ~.
\end{equation}
This condition, which is necessary so that terms do not appear in the potential which are parametrically larger than those responsible for the large volume vacuum, turns out to be extremely restrictive.

Crucially, in describing the Swiss Cheese condition, we have assumed a partition of the geometry into large and small four-cycles.
For an arbitrary geometry, the basis for $\Div(\cX)$ that is natural given how the space was constructed may not be compatible with such a partition even if a large volume vacuum exists.
In an arbitrary divisor basis, the large and small cycles generically mix together.
In performing an algorithmic scan for Swiss Cheese manifolds, it is important to include an initially arbitrary basis transformation that yields a partition into small and large four-cycles.

We define the rotation
\begin{equation}\label{eq_4cyclebasech}
\tau_i = \sum_{\tilde\jmath=1}^{h^{1,1}} A_{i}{}^{\tilde\jmath} \tilde\tau_{\tilde\jmath}
\end{equation}
for some non-degenerate matrix $A\in GL(h^{1,1};\IZ)$.
In what follows, we will search for suitable solutions for the entries of $A$ such that the divisors $\tau_i$ are separated out into the two desired classes:
\begin{equation}\label{eq_indexsplitting}
\bal
& \text{large cycles $\tau_{{\rm L}_A}$:} && \tau_I ~, && \text{$I=1,\ldots,N_{\rm large}$} ~, \\
& \text{small cycles $\tau_{{\rm s}_a}$:} && \tau_\alpha ~, && \text{$\alpha=N_{\rm large}+1,\ldots,h^{1,1}(\cX)$} ~. \\
\eal
\end{equation}
The problem of identifying Swiss Cheese geometries reduces in essence to characterizing the LVL in an arbitrary basis, determining whether it exists, and checking that the inverse K\"ahler potential has the correct scaling properties. \\

\vspace{-0.4cm}

\sect{Rewriting the Swiss Cheese Condition}
In order to deduce whether $\cX$ is Swiss Cheese, one could simply solve for the volume of ${\cal X}$ as a function of the $\tau$ and check the scaling of the inverse K\"ahler potential directly.
However, this naive procedure turns out to be extremely inefficient computationally in all but the very simplest of cases.
Instead, we reformulate the conditions for a large volume vacuum in terms of the K\"ahler parameters $t$.

Restricted to each divisor four-cycle $D_i$, the intersection form reduces to a symmetric matrix $(\kappa_{(i)})_{jk} \ce \kappa_{ijk}$.
Furthermore, let $\vec t = (t^1,\dots,t^n)$ denote a (column) vector of the K\"ahler parameters with respect to the expansion of the symplectic form in \eqref{eq:CY3SymplecticJ}.
The four-cycle volumes \eqref{eq_taudef} can then be rewritten as $\tau_i = \frac{1}{2}\kappa_{ijk} t^j t^k = \frac{1}{2} \vec t^* \kappa_{(i)} \vec t$, where $\vec t^*$ refers to the transposed row vector of $\vec t$.
Due to the correspondence between the four-cycle volumes $\tau_i$ and the K\"{a}hler parameters $t^i$, the LVL sends particular linear combinations of the K\"ahler parameters $t^i$ to infinity.
We split the K\"ahler parameter vector into the form
\begin{equation} \label{tvec}
\vec t = \lambda_A \vec t_{{\rm L}_A} + \gamma_a\vec t_{{\rm s}_a} ~,
\end{equation}
where $\lambda_A$ and $\gamma_a$ are positive real numbers.
The (potentially different) large volume limits correspond to the limits $\lambda_A\to\infty$ for some or several $A=1,\ldots,N_{\rm large}$.
Therefore, $\vec t_{{\rm L}_A}$ and $\vec t_{{\rm s}_a}$ for $a=1,\dots,N_{\rm small}$ refer to the large and small directions in the K\"{a}hler parameter space.
After inserting this splitting into \eqref{eq_taudef}, we obtain
\begin{eqnarray}
\tau_i &=& \frac{1}{2}\left[ \lambda_A\lambda_B \cdot (\vec t_{{\rm L}_A}^* \kappa_{(i)} \vec t_{{\rm L}_B}) + 2\lambda_A \gamma_b \cdot (\vec t_{{\rm L}_A}^* \kappa_{(i)} \vec t_{{\rm s}_b})\right. \nonumber \\
&& \left. {}+\gamma_a \gamma_b \cdot (\vec t_{{\rm s}_a}^* \kappa_{(i)} \vec t_{{\rm s}_b}) \right] ~.
\end{eqnarray}
Note that the first two terms in this expansion contain powers of the associated large direction parameters $\lambda_A$, whereas the last term is independent of them.

Due to the general basis change \eqref{eq_4cyclebasech}, we can pick any $N_{\textnormal{large}}$ number of $\tau$s to correspond to our large four-cycles $\tau_I$ with the remaining K\"ahler moduli corresponding to small cycles $\tau_{\alpha}$.
From the $\lambda$ power counting in the expansion, the scaling of the large cycles and small cycles then demands
\begin{equation}
\bal
& \text{large cycles $\tau_I$:} && \vec t_{{\rm L}_A}^* \kappa_{(I)} \vec t_{{\rm L}_B} \not= 0 ~~\text{OR}~~ \vec t_{{\rm L}_A}^* \kappa_{(I)} \vec t_{{\rm s}_b}\not= 0 ~, \\
& \text{small cycles $\tau_\alpha$:} && \kappa_{(\alpha)} \vec t_{{\rm L}_A} = 0
\eal
\label{eq:conds0}
\end{equation}
for the respective divisors, such that as each $\lambda_A\to\infty$, one combination of the large cycle volumes $\tau_I\to\infty$.

Aside from the distinction between large and small cycles, there are also the conditions \eqref{eq_invKaehlerDiag} on the inverse K\"ahler metric.
In the expansion in inverse volume, for small cycles $\alpha$,
\begin{equation}
\frac{K^{-1}_{\alpha\alpha}}{\cV} = -\frac{4}{9} \kappa_{\alpha\alpha i}t^i + \frac{4(\tau_\alpha)^2}{\cV} + \cO(\cV^{-2}) ~;
\label{eq:this}
\end{equation}
the second term goes to zero in the LVL by construction as the small cycle volumes $\tau_\alpha$ remain finite when $\cV\to\infty$.
In terms of the matrix/vector notation for the restricted intersection matrices on the divisors, \eqref{eq:this} in the limit asserts that
\begin{equation}
\frac{K^{-1}_{\alpha\alpha}}{\cV} = -\frac{4}{9} (\kappa_{(\alpha)}\vec t)_\alpha
\label{eq:lhsrhs}
\end{equation}
and, because of \eqref{eq_invKaehlerDiag}, it must scale as $\sqrt{\tau_\alpha}$.
Since this scaling only involves small cycles, the large volume direction has to vanish on the right hand side of \eqref{eq:lhsrhs}, leading to the condition
\begin{equation}\label{eq_invKahlerCond}
(\kappa_{(\alpha)}\vec t_{{\rm L}_A})_\alpha = 0 ~.
\end{equation}
By \eqref{eq:conds0}, this requirement is automatically satisfied for all $\alpha$.

We can express the non-triviality and non-colinearity of the vectors $\vec{t_{L_A}}$ and $\vec{t_{s_\alpha}}$ by requiring
\begin{equation}
\det\Big( \vec t_{{\rm L}_1},\dots,\vec t_{{\rm L}_{N_{\rm large}}},\vec t_{{\rm s}_1},\dots,\vec t_{{\rm s}_{N_{\rm small}}} \Big) \not= 0 ~.
\label{eq:dets}
\end{equation}
To establish the possibility of a large volume vacuum, it suffices to check whether a solution to all of the conditions we have described exists.\\

\vspace{-0.4cm}

\sect{The Algorithm}
The input data for our algorithm are the triple intersection numbers and a description of the K\"ahler cone of the Calabi--Yau to be considered.
Since these data will not necessarily be provided in a basis compatible with the large and small cycle structure of the LVL, we consider the associated basis of four-cycle volumes to be the tilded one given in equation \eqref{eq_4cyclebasech}.

Since $\cX$ has non-degenerate intersection numbers, by Poincar\'e duality we may invert \eqref{eq_taudef}, which relates the $t^i$ and the $\tau_i$.
This allows us to eliminate the two large cycle conditions, as their existence follows from the non-colinearity condition as the complement of the small cycles.
Combining all we have learned, this then leaves us with the following minimal set of conditions to find a Calabi--Yau threefold capable of admitting a LVS vacuum:
\begin{equation}\label{eq_redCondsh11_2}
\boxed{
\bal
& \text{small cycles:} & A_{\alpha}{}^{\tilde\jmath} ( \kappa_{(\tilde\jmath)} \vec t_{{\rm L}_A} ) {}&{} = 0 ~, \\
& \text{basis change:} & \det [A_{i}{}^{\tilde\jmath}] {}&{} \not= 0 ~, \\
& \text{$K^{-1}$ scaling:} & A_{\alpha}{}^{\tilde\imath} A_{\alpha}{}^{\tilde\jmath} (\kappa_{(\tilde\imath)} \vec t_{{\rm s}_a})_{\tilde\jmath} {}&{}\not= 0 ~, \\
& \text{non-triviality:} & \det\big( \vec t_{{\rm L}_1},\dots, \vec t_{{\rm s}_1},\dots \big) {}&{} \not= 0 ~, \\
& \text{K\"ahler cone:} & \lambda_A(\vec t_{{\rm L}_A})^{\tilde\imath} \,+\, \gamma_a (\vec t_{{\rm s}_a})^{\tilde\imath} {}&{} \ge 0 ~. \;
\eal
}
\end{equation}

We must solve these for $A_{i}{}^{\tilde\jmath}$, $\lambda_A$, $\gamma_a$, $\vec t_{{\rm L}_A}$, and $\vec t_{{\rm s}_a}$.
We point out that the last condition, which checks that the conditions can be solved in the interior of the K\"ahler cone, states that this cone is simply the positive orthant.
In the cases we shall study here this has been ensured by an additional transformation on the triple intersection form, and we have restricted ourselves to cases with simplicial K\"ahler cones \cite{longpaper}.
In general a more complicated description of this condition must be input.

The system \eqref{eq_redCondsh11_2} contains many redundancies in the variables for which we are solving.
Given their definition \eqref{tvec}, we are only concerned about the direction of the vectors $\vec t$ in the K\"{a}hler moduli space, such that we can use the $h^{1,1}$ redundancies of their lengths $|\vec t|$ to fix some of the inequalities in \eqref{eq_redCondsh11_2} to equalities.
Furthermore, the basis change matrix $A_{i}{}^{\tilde\jmath}$ is not required to be an arbitrary $GL(h^{1,1}; \IZ)$ matrix in order to isolate the small cycles from the large cycles, and some of the residual freedom can be used to fix some of the components of $A$.
By these means, the second and fourth inequalities in \eqref{eq_redCondsh11_2} can be set to plus or minus unity.

In solving the resulting equations, or in proving that they do not admit a solution, we take a two step approach.
Firstly, we analyze the first, second, and fourth equations of our system, which, after the redundancy fixing above, describe an algebraic variety.
Using methods of computational algebraic geometry and in particular the program {\tt Singular}~\cite{DGPS}, we check the complex dimension of the solution space of these equations using a Gr\"obner basis computation~\cite{SV}.
If the associated dimension is minus one, i.e., if the equations describe the empty variety, there are, in particular, no real solutions for the variables and the case of interest admits no large volume vacuum.
If the dimension of the ideal is greater than or equal to zero, we must solve the associated equations over the reals.
To facilitate this solution, which occurs in the second of our two steps, we primary decompose the ideal using the GTZ algorithm~\cite{GTZ}.
This returns sets of simpler equation sets, one for each of the irreducible solution spaces of the system.

Secondly, we proceed to search for a solution to the simplified equation system, if one exists, with the remaining inequalities in \eqref{eq_redCondsh11_2} added back in.
This analysis is performed using standard techniques available in packages such as {\tt Mathematica}.
The simplification afforded in these cases by the primary decomposition of the initial equation set is enough to allow the computation to finish in reasonable time.
In fact, we find that cases which are not ruled out by the dimension check in the first step of our analysis are almost always Swiss Cheese.

The output consists of a Boolean determining whether the manifold is Swiss Cheese and, in the case of a positive result, a matrix $A$ explicitly identifying the large and small cycles in terms of the original basis of four-cycles.\\

\vspace{-0.4cm}

\sect{Results}
A scan over the Calabi--Yau manifolds defined as complete intersections in the products of projective spaces (CICYs) showed that there are no Swiss Cheese geometries of this class for $h^{1,1}\le 4$.
Because CICYs at low $h^{1,1}$ are all favorable, they lack the blowup cycles that could be associated with the small cycles in the LVL, and thus this is not a surprise.
Nevertheless, this class, along with the known Swiss Cheese manifolds, provides a useful test of the algorithm.

Implementing the algorithm on the Kreuzer--Skarke dataset of hypersurfaces in toric ambient spaces for $h^{1,1} \leq 4$ results in $418$ Swiss Cheese Calabi--Yau manifolds with one large cycle.
The results are summarized in the following table.

\vspace{-0.0cm}
\begin{table}[h!]
\begin{center}
\begin{tabular}{| c | c | c |c|}\hline
$h^{1,1}$ & \# of Cases Scanned & \# of Swiss Cheese & \# of Strong Cheese \\ \hline\hline
$2$ & $39$ & $22$ & $22$ \\ \hline
$3$ & $266$ & $94$ & $50$\\ \hline
$4$ & $3513$ & $302$ & $106$ \\ \hline
\end{tabular}
\end{center}
\end{table}

\vspace{-0.4cm}
The Kreuzer--Skarke database contains, respectively, $36$, $244$, and $1197$ polytopes whose resulting manifolds have $h^{1,1}=2$, $3$, and~$4$.
In many cases there are many possible triangulations for each polytope, and thus the number of geometries to consider are $39$, $306$, and $5930$ for $h^{1,1}=2$, $3$, and~$4$, respectively.
Of these, the above table counts those whose K\"ahler cones are simplicial.
We note that while the overall volume for all $h^{1,1}=2$ Swiss Cheeses can always be recast in the Strong Cheese form in~(\ref{eq_SwissCheeseScaling}), this can only be done for $50$ of the $h^{1,1}=3$ cases and $106$ of the $h^{1,1}=4$ cases.

One can ask how far in $h^{1,1}$ it will be possible to push these scans.
In particular, the Gr\"{o}bner basis computation performed by {\tt Singular} is a highly optimized implementation of the Buchberger algorithm.
This has a worst case scenario double exponential scaling behavior in the number of unknown variables~\cite{MM}.
Solving \eqref{eq_redCondsh11_2} for $h^{1,1}\le 4$ can be done in a matter of seconds or minutes, and for the CICYs it has been checked that scans up to $h^{1,1} = 8$ can easily be finished on a standard desktop machine.
At this stage in our analysis, however, the full possibility of removing redundancies from the variables of equation system \eqref{eq_redCondsh11_2} has not been utilized.
At present, it is not clear how far beyond $h^{1,1}=8$ it will be possible to push the algorithm once the potentially double exponential improvement in calculation speed afforded by removing additional redundancies is incorporated.
The results of this work will be presented in a forthcoming publication~\cite{longpaper}. \\

\vspace{-0.4cm}

\sect{An Example and Future Work}
To provide a concrete example of a Swiss Cheese Calabi--Yau found by this algorithm let us consider a case where $h^{1,1}=4$.
The intersection form for this case gives the following expression for the volume \eqref{eq:InnerVolume}:
\begin{eqnarray}\nonumber
6 \cV &=& (2 t_1^3+ 3t_1^2(6 t_2+t_3+t_4)+ 9 t_1(4 t_2^2-t_3^2-t_4^2 \\ \nonumber && +4 t_2(t_3+t_4))+ 3(8t_2^3-5 t_3^3-6t_3^2t_4 -6 t_3 t_4^2 \\ \label{eq_h11twoEx1} && -5 t_4^3 +12 t_2^2(t_3+t_4) + 6 t_2(t_3+t_4)^2)) ~.
\end{eqnarray}
The scanning algorithm provides the base change matrix $A_{i}{}^{\tilde\jmath}$, such that by a rotation of the four-cycles $\tau_i = A_i{}^{\tilde\jmath} \tilde\tau_{\tilde\jmath}$, we obtain
{\small
\begin{equation}\label{eq_h11twoEx1b} \nonumber
\left( \begin{array}{cccc} 0 & 1 & 0 & 0 \\ -3 & 1 & 0 & 1 \\ 6 & -1 & -2 & -2 \\ -3 & 1 & 1 & 0 \end{array} \right)
\left( \begin{array}{c} \tilde\tau_1 \\ \tilde\tau_2 \\ \tilde\tau_3 \\ \tilde\tau_4 \end{array} \right) =
\left( \begin{array}{c} 3(t_1 + 2 t_2 + t_3 +t_4)^2 \\ \frac{1}{2}(t_1 + 3t_3)^2 \\ (t_1+3(t_3+t_4))^2 \\ \frac{1}{2}(t_1+3 t_4)^2 \end{array} \right) ~.
\end{equation}}
In this four-cycle basis, the volume \eqref{eq_h11twoEx1} takes the Strong Cheese form described in \eqref{eq_SwissCheeseScaling}:
\begin{equation}
\cV = \frac{1}{18}\left( \sqrt{3} \tau_{\rm L}^{\frac{3}{2}} - 2 \sqrt{2} \tau_{\rm s ,1}^{\frac{3}{2}} -\tau_{\rm s ,2}^{\frac{3}{2}} - 2 \sqrt{2} \tau_{\rm s ,3}^{\frac{3}{2}} \right) ~.
\label{h2vol}
\end{equation}

Our initial scan of smooth compactification manifolds with small numbers of K\"{a}hler parameters $h^{1,1}\le 4$ shows that the Calabi--Yau threefold landscape is richly populated by Swiss Cheese geometries.
This is fortuitous as many more constraints must be imposed upon a Calabi--Yau threefold than those considered here if it is to give rise to a phenomenologically acceptable vacuum (as emphasized from a scanning perspective in~\cite{Cicoli:2012vw}).
In a future publication~\cite{longpaper}, we will present the results of performing the scan outlined here over as large a set of Calabi--Yau threefolds as possible.
Interestingly, requiring the existence of a LVS already constrains the space of allowed intersection numbers significantly~\cite{longpaper}.
The future of this research program will then consist of cataloguing ever more detailed properties of these geometries in a systematic way.
Initial steps in this regard will be to catalogue which structures are available for both moduli stabilization and model building on each manifold, and thus which variants of the LVS can be realized in each case.
The database will be made freely available in a standardized format so that our results may be exploited and supplemented by other groups.\\

\vspace{-0.4cm}

\sect{Acknowledgements}
We thank L.~Anderson, R.~Blumenhagen, M.~Cicoli, T.~Grimm, S.~Krippendorf, C.~Mayrhofer, and F.~Quevedo for useful discussions.
This work was partially supported by the NFS-Microsoft grant NSF/CCF-1048082, the NSF under grant PHY05-51164, EPSRC under grant EP/G007985/1, and the DST/NRF SARChI program.


\vspace{-0.6cm}

\begin{thebibliography}{99}
\vspace{-0.6cm}
\bibitem{Balasubramanian:2005zx}
V.~Balasubramanian, P.~Berglund, J.~P.~Conlon, and F.~Quevedo,
JHEP {\bf 0503}, 007 (2005)
[hep-th/0502058].
\bibitem{Giddings:2001yu}
S.~B.~Giddings, S.~Kachru, and J.~Polchinski,
Phys.\ Rev.\ D {\bf 66}, 106006 (2002)
[hep-th/0105097].
\bibitem{Kachru:2003aw}
S.~Kachru, R.~Kallosh, A.~D.~Linde, and S.~P.~Trivedi,
Phys.\ Rev.\ D {\bf 68}, 046005 (2003)
[hep-th/0301240].
\bibitem{Cicoli:2008va}
M.~Cicoli, J.~P.~Conlon, and F.~Quevedo,
JHEP {\bf 0810}, 105 (2008)
[arXiv:0805.1029 [hep-th]].
\bibitem{pmh}
P.~M.~H.~Wilson,
Invent.~math.~98, 139-155 (1989).
\bibitem{Blumenhagen:2008zz} 
  R.~Blumenhagen, V.~Braun, T.~W.~Grimm and T.~Weigand,
  Nucl.\ Phys.\ B {\bf 815}, 1 (2009)
  [arXiv:0811.2936 [hep-th]].
\bibitem{Cicoli:2011it}
M.~Cicoli, M.~Kreuzer, and C.~Mayrhofer,
JHEP {\bf 1202}, 002 (2012)
[arXiv:1107.0383 [hep-th]].
\bibitem{Cicoli:2012vw}
M.~Cicoli, S.~Krippendorf, C.~Mayrhofer, F.~Quevedo, and R.~Valandro,
arXiv:1206.5237 [hep-th].
\bibitem{DGPS} {\sc Singular} 3.1.3
W.~Decker, G.-M.~Greuel, G.~Pfister, and H.~Sch{\"o}nemann, 2011 (http://www.singular.uni-kl.de).
\bibitem{SV}
For related work applying such methods to moduli stabilization see,
J.~Gray, Y.~-H.~He, and A.~Lukas,
JHEP {\bf 0609}, 031 (2006)
[hep-th/0606122];
J.~Gray, Y.~-H.~He, A.~Ilderton, and A.~Lukas,
Comput.\ Phys.\ Commun.\ {\bf 180}, 107 (2009)
[arXiv:0801.1508 [hep-th]]; and
J.~Gray,
Adv.\ High Energy Phys.\ {\bf 2011}, 217035 (2011)
[arXiv:0901.1662 [hep-th]].

\bibitem{GTZ}
P.~Gianni, B.~Trager, and G.~Zacharias,
J.~Symb.~Comp. 6, 149-167 (1988).
\bibitem{MM}
H.~M.~M\"oller and F.~Mora,
Comput.~Sci. 174, 172-183 (1984).
\bibitem{longpaper}
J.~Gray, Y.-H.~He, V.~Jejjala, B.~Jurke, B.~Nelson, and J.~Sim\'on,
to appear.
\end{thebibliography}
\end{document}